ON THE EXPRESSIVENESS OF LINE DRAWINGS

Harm Hollestelle, Amsterdam, The Netherlands
This study was made possible with support of The Netherlands Foundation for Visual Arts, Design and Architecture

ABSTRACT

Can expressiveness of a drawing be traced with a computer? In this study a neural network (perceptron) and a support vector machine are used to classify line drawings. To do this the line drawings are attributed values according to a kinematic model and a diffusion model for the lines they consist of. The values for both models are related to looking times. Extreme values according to these models, that is both extremely short and extremely long looking times, are interpreted as indicating expressiveness. The results strongly indicate that expressiveness in this sense can be detected, at least with a neural network.

1. INTRODUCTION

Images are increasingly studied in order to create methods to select information from them. For instance one tries to recapture three dimensional shape from two dimensional pictures or photographs [ref.1]. Also the perception of asymmetric or disorderly deviations from a regular pattern has been investigated [ref.2, 3, 4, 5]. In these studies a measure of perceptibility in the sense of  accurateness is derived. Expressiveness of drawings has been discussed systematically in relation to the psychological characterisation of children [ref.6]. This discussion is mainly qualitative and the evaluation of drawings remains personal and descriptive.
There is a long and respected history in the field of perception for examining line drawings, particularly in order to derive information about human vision in general. Likewise, there is a strong, although not quite as long, history of studying art images in perception and for studying the effectiveness of visualizations in Computer Science. Finally, neural networks and the automated analysis of images have received an extremely large amount of attention over the last 3 decades. In particular, there have been several recent algorithms (in the fields of visualization, visual analytics and computational aesthetics) that automatically examine line drawings for their artistic (that is sculptural, 3D) aspects [ref.7] and informational aspects. However these algorithms are not the subject of investigation in this article. Here the discussion centers on expressiveness of line drawings. With this is meant how much the drawing reaches out to you. This is a qualitatively different kind of information, or, better, it is a feeling. It is independent of human vision or perception accurateness, visualization, information transfer or 3D evaluation.

The group of line drawings investigated here are made, and appreciated, by student physicists for practicum reports. They are of a type changing from portrait to landscape metaphorically. When considering automation of image recognition one generally means the recognition of some features among many other, from faces on a realistic photograph [ref.8] or from an aerial survey of, for instance, farm land [ref.9]. These kinds of image recognition are the same as the problem presented here in that only a few characteristic features of the picture is focused upon. However they are not the same with respect to the fact that here the uncertainty in the results is much larger. This is due to the fact that the set of drawings is described by aspects that can conflict with, and are complementary to, each other. Because of this the learning set for the perceptron never reaches near zero average deviations from targets after iteration. Possibly conflicting expressiveness aspects are the necessary result of the fact that they are independent and differ inherently from each other and that used as learning set are drawings with extreme values of these aspects.

Student drawings provide a large set of comparable drawings with similar content. Recent drawings of students are studied for a variety of reasons, as there is for psychological evaluation [ref.10], to improve on the visualisation of data results [ref.11] and to monitor the improvement of understanding of parts of science theory [ref.12].

Studied here is a set of physics student drawings in order to define measures for expressiveness. By counting the elementary parts of these drawings, that are all black and white line-drawings, one can quantitatively discuss expressiveness aspects of them. In this way expressiveness becomes accessible for automation with a neural network. Elementary parts, building blocks like soft curves and corners, of drawings were discussed for the first time by Kellogg [ref.13], in the case of children's drawings. Also the metaphorical use of elementary parts, that is in this case curves and corners, in drawings by infants has been described and interpreted before [Adi-Japha, ref.14]. This last study provides a modelling of soft curves in lines that is the starting point for the present investigation.

2. SUMMARY

Expressiveness of a drawing can be related to the time or duration a drawing captures ones attention. When a drawing makes one look at it attentively for extreme values in duration in time, extreme values for expressiveness can be attributed to it as well. With extreme values is meant both an extremely short or an extremely long duration in time. When the drawing shouts out at you, you can react in two ways: you can run away hastily after a short moment, or you can stay, paralyzed so to speak, for a long time.

In this study two different models are used to derive time scales, in terms of the above mentioned elementary aspects, for the attention a line drawing demands when looked at. One is a model using concepts of diffusion theory to describe the drawing as a cluster of lines through which ones attention diffuses locally. The other is a kinematical model that describes the flow of attention following the lines in a sequel throughout the drawing on average. A third aspect is related to how near the drawing literally seems to be to the observer. The first aspect is dependent on small scale character of the drawing, the second one on average large scale character. The third aspect is an out of the plane aspect. In this way a reasonably complete description, with respect to time and space, of expressiveness is given. This is described in paragraph 3.

The character of a perceptron (neural network) is such that it fits the study of expressiveness aspects. Some of the elementary aspects, necessary for the modeling above, are interpretation dependent and cannot easily be measured directly. A perceptron can replace the laborious measurements of these aspects, once they are done for a specific learning set of drawings. Used for training the perceptron is a learning set of line drawings with extreme values for the expressiveness aspects. After having learned from these drawings, the perceptron provides, for all other drawings, values for the interpretation dependent elementary aspects. These perceptron values then can readily be applied to calculate the values for expressiveness according to the models above.

15 drawings were selected for the learning set and 15 drawings were used for testing the perceptron. The perceptron is a three layer neural network with 12 input, 5 intermediate (hidden) and 5 output cells. As reception function used is the Fermi-(sigmoid-)function. The learning set of 15 drawings is a very small one. It is limited in size by the capacity of the computer used. Nevertheless, the results are such that the perceptron seems to perform well with this small learning set. Also 12 input values that should describe the characteristics of a line drawing is a small number. Bear in mind however that it is not the complete information transfer that is intended but instead the evaluation of measures of expressiveness. Measures of expressiveness that give values to something that is intrinsically a qualitative phenomenon.

It turns out that the perceptron, for the test set of drawings, can reproduce the values for the three aspects of expressiveness as they are derived by hand. This ensures that the values are attributed to these aspects systematically. A 1-norm soft margin support vector machine (SVM) was used to repeat this investigation. The results for the SVM however are much less clear. This, together with the results for the perceptron, is discussed in paragraph 4.

The resulting time scales for the two models mentioned above can be applied to describe qualitatively results derived from time measurements of infant attention to faces. For the description of time scales of infant attention to faces spectral functions of the source, that is the face and modifications of it, have been used [ref.15]. However spectral functions do not seem to give the correct measures for infant attention in terms of time scales [ref.16]. Johnson [ref.16] measured time scales of infant attention for both moving faces and still faces. These results can be interpreted in terms of the two time scales of the two models presented above (paragraph 5).

3. EXPRESSIVENESS ASPECTS IN TERMS OF LOOKING TIMES

In this research 30 recent drawings made by physics students of Utrecht University for practicum reports are investigated. Making these drawings is part of the standard curriculum, they are not specially made for this investigation. All drawings are black and white line drawings. For each of these drawings several elementary aspects, building blocks, were measured or counted. The elementary aspects involved are listed below. They include for instance the soft curves and corner bends present in a drawing. From the values for these elementary aspects, values for three different aspects of expressiveness are derived. The expressiveness aspects are called (a), (b) and (c) respectively for convenience. Both aspect (a) and (b) are looking time scale measures. They are defined by a diffusion model for aspect (a) and a kinematical model for aspect (b). Aspect (a) is an average measure for the time scale to transverse with one's attention a standard distance locally in the drawing. One's attention is described as diffusing through the cluster of line segments the drawing consists of. Therefore this is a small scale aspect. Aspect (b) is an average measure for the time scale one needs to follow with one's attention a sequel of line segments throughout the drawing, like a melody. This is a large scale aspect. It depends on approximating and generalizing the time needed to follow one soft bend or corner bend in an average line segment. The modeling of the kinematics for soft bends was already given in [Adi-Japha, ref.11]. Aspect (c) describes how near a drawing literally seems to be. This is an out of the plane aspect. Thus the aspects (a), (b) and (c) are inherently different from and complementary to each other and form a complete set of aspects describing expressiveness in all three dimensions.

The details of the models that are used will be described elsewhere [ref.17]. The relations, derived from the models, between the expressiveness aspects and the elementary aspects are:
For aspect (a): value(a) = (l-div) / (l-segm) where (l-div) is a measure for the line-diversity of the drawing and (l-segm) is the average line segment length for the drawing.
For aspect (b): value(b) = (#e) (( %s) (l-segm)$^{2/3}$ + (%c) (l-segm)$^{2/5}$) where (#e) is the total number of bends and ends of all line segments together of the drawing and (%s) is the percentage of soft bends, curves, and (%c) is the percentage of corner bends.
Aspect (c) is defined as: value(c) = (M) (l-div) / (S) where (M) is the length of the main independent form in the drawing and (S) is the smallest distance in the drawing. The average line segment length of a drawing can be approximated by (l-segm) = ((S)(L))$^{1/2}$, taking the root of the product of the smallest distance (S) and the largest distance (L). Among the elementary aspects that are present in these definitions some, that is (L), (S), (l-div), can be counted or measured directly from each real drawing. The other aspects however ((#e), (%s), (%c), (M)) are subject to interpretation of how line segments do intersect and how this is counted.

By determining the values for the elementary aspects by hand for all the drawings in the studied set one can investigate as to whether there are correlations between them. This has to be done with consideration for the fact that this set of drawings is a specific one. Report drawings of physicists are special because they are made by the same person who built the experimental set-up of the experiment. Also they are special because they are made with the intention to explore the experimental question the set-up was intended to answer in the first place and to explore physical questions about space, time and matter in general. As will be argued in the discussion section, paragraph 6, these drawings still can stand as example for all line drawings when studying aspects of expressiveness.

The correlations that were found are trends rather than exact relations. The following trends can be distinguished using the abbreviations introduced above. For the elementary aspects: (%s) tends to decrease and (%c) tends to increase for increasing (#e). (S) tends to decrease and (L) tends to increase for increasing (#e). For the expressiveness aspects there is found: (a) and (c )tend to correlate with 1/(S) while (b) tends to correlate with (L).
This suggests the correlation trend that (S) decreases with increasing (L) and thus (a) increases with increasing (b). Decreasing (S) for increasing (L) seems counter intuitive, however notice that for larger drawings the artist finds more space, and time, to concentrate safely on detail. This seems to be the case at least for this set of drawings.
Aspect (a) is a small scale aspect intuitively related to a function of (S) rather than to a function of (L) and aspect (b) is a large scale aspect intuitively related to (L) rather than to (S). So these two aspects seem to be independent. When (S) decreases with increasing (L) one finds that (a) increases with increasing (b). So the positive correlation trend between these two looking times aspects (a) and (b) depends on the correlation trend between 1/(S) and (L) and is really new.

4. AUTOMATION

The neural network used for this study is a three layer perceptron, with the Fermi-function as reception function and with twelve input cells, five hidden cells and five output cells. Applied is the algorithm as in [ref.18]. As input values are chosen parameters and elementary aspects that can be measured by a computer in a direct way. The output values are those belonging to the elementary aspects that are subject to interpretation. The output values combined with the input values are sufficient to determine the perceptron values for the expressiveness aspects, P(a), P(b) and P(c ).
The first eight input values are the result of reducing the line drawing composition to eight values. To do this, each line drawing is interpreted as a configuration of points for each intersection of lines, or each bend, or end in a line. This point configuration is then reduced to 4 values that are the averages for the point co-ordinates above and below and left and right of the centre point of the drawing, plus 4 values that are the average deviations thereof. The drawing can now be visualized by the two rectangles around the centre point, that constitute the inner or outer border respectively of the averages, minus or plus their average deviations.  Because the rectangles can be found placed (a)symmetrically with respect to each other this reduction preserves, to some extent, the (a)symmetry and composition of the line drawing.
To these 8 values are added the number of points (#p), the large scale aspect (L), the small scale aspect (S) and the line diversity (l-div). These 12 values are the input values for each drawing separately. All values are measured or counted in millimeters or numbers except (S) that is measured in 0.1mm. (#p), (L) and (S) can directly be measured by a computer, however (l-div) is an aspect that has to be given beforehand. It does not depend on the composition or sizes of the drawing but instead on simple and straightforward visual information. It can be compared to content, that is whether a drawing has theoretical content or experimental set-up content or both. Together with content, (l-div) in this study is considered as a generalized aspect depending rather on general evaluation then on measurable quantities. Still (l-div) is added as input value because it is needed for the derivation of the expressiveness aspects.

The 5 output values are respectively the values ((#p)/(#e)), (%s), (%c), the composition type of the drawing that is changing from portrait to landscape (values ranging from 0.2 to 0.8) and the value ((M)/(L)). Given the input values, from these output values the perceptron values for the expressiveness aspects can be derived. For both the learning set and the test set of 15 drawings each, all input values and output values were also measured by hand.

For the learning set an average deviation of targets was reached, after 5000 iterations, of approximately 0.12 for target output values ranging from 0.0 to 1.0. Started was with a specific choice of weights whereas a random choice of starting weights did not decrease the resulting deviation. Also using another number (10, 15) of hidden cells did not improve the results. Starting weights between input cells and inner cells were all positive and the same for each inner cell, all other starting weights were equal to +1 or -1. Different starting weights for different input cells were used because also the average input values for these cells differed. Otherwise the input cells with the largest input values like (#p) (nr 9) and (L) (nr 10) would dominate. The starting values and final values after 5000 iterations are listed below:

Starting weight diagram: 12 input cells / 5 inner cells, weights x 1000

|        | 1  | 2  | 3  | 4  | 5  | 6  | 7  | 8  | 9 | 10 | 11 | 12 |
|--------|----|----|----|----|----|----|----|----|---|----|----|----|
| 1 to 5 | 50 | 50 | 50 | 50 | 50 | 50 | 50 | 50 | 1 | 1  | 10 | 10 |

Starting weight diagram: 5 inner cell / 5 output cells

|              | Inner cell 1 | 2  | 3  | 4  | 5  |
|--------------|--------------|----|----|----|----|
| Output cell 1 | 1            | 1  | 1  | 1  | -1 |
| 2            | 1            | 1  | 1  | -1 | 1  |
| 3            | 1            | 1  | 1  | -1 | -1 |
| 4            | 1            | 1  | -1 | -1 | -1 |
| 5            | 1            | -1 | 1  | -1 | -1 |

Final weight diagram: 12 input cells / 5 inner cells, weights x 1000.

|   | 1  | 2   | 3   | 4   | 5   | 6   | 7  | 8  | 9   | 10  | 11   | 12  |
|---|----|-----|-----|-----|-----|-----|----|----|-----|-----|------|-----|
| 1 | 83 | 327 | 47  | -36 | 185 | 158 | 38 | 34 | 12  | -21 | -158 | -2  |
| 2 | 47 | 43  | 55  | 46  | 52  | 50  | 51 | 50 | 10  | -53 | -3   | 21  |
| 3 | 86 | 81  | 291 | -63 | -22 | 9   | 10 | 27 | -53 | -13 | -37  | 158 |
| 4 | 19 | 69  | 92  | 96  | -7  | 36  | 57 | 58 | 54  | -10 | -114 | 64  |
| 5 | 72 | -16 | 48  | 139 | 36  | 32  | 56 | 65 | -12 | -16 | 2    | 226 |

Final weight diagram: 5 inner cells / 5 output cells, weights x 1000.

|              | Inner cell 1 | 2     | 3     | 4     | 5     |
|--------------|--------------|-------|-------|-------|-------|
| Output cell 1| 1737         | 728   | 912   | 3656  | -808  |
| 2            | 867          | 836   | -3168 | -665  | 1831  |
| 3            | 557          | 1062  | 1093  | -1139 | -2168 |
| 4            | 3720         | 1136  | -1904 | -535  | 4     |
| 5            | 1816         | -1198 | 5837  | 228   | -2136 |

For the test set the resulting average deviation is approximately 0.17. These results are very reasonable when one considers that the output values are measures of aspects that are only vaguely described. When one calculates the values P(a), P(b) and P(c ) for the three expressiveness aspects from the output values of the perceptron one finds the relations P(a) – (a), P(b) – (b) and P(c ) – (c ) to be linear in an acceptable way (figures 1 to 3). Thus the perceptron can reproduce the values measured by hand for these aspects and also confirms the correlation trends for these aspects described in paragraph 2. For convenience the short hand notation (a) is used for the value of aspect (a) measured by hand and so forth.

The drawings investigated with the perceptron were also analysed with a support vector machine (SVM). Following the standard algorithm for a 1-norm soft margin SVM [ref.19] with as kernel (K)the usual vector product minus a parameter $\underline{b}$, one derives for a learning set of drawings a weight vector $\underline{w}$ that can discriminate between classes of drawings having large resp. small values for a certain property. For the aspects (a), (b) and (c ) weight vectors were derived using 10 drawings as learning set which was the maximum for the present computer used. The input values $\underline{x}$ for each drawing were the same as used for the perceptron. The weight vectors were tested with the 20 remaining drawings. For each aspect the weight vector $\underline{w}$ and the parameter $\underline{b}$ were chosen in such a manner that for the learning set of drawings the kernel $K(\underline{w},\underline{x}) = <\underline{w},\underline{x}>-\underline{b}$ is positive for drawings with large values for the aspect resp. negative for drawings with low values.
Using these kernels one can plot the resulting SVM values SVM(a), SVM(b) and SVM(c )for the test drawings against the values for each drawing measured by hand. The resulting relation SVM(b) – (b) turns out to have an increasing line as an average, however with large deviations. Thus the SVM reproduces to some extent for aspect (b) the results derived by hand. However for aspects (a) and (c) this confirmation is not found since no increasing linear average is found for the relations SVM(a) – (a) an SVM(c ) – (c ) (figures 4 to 6).

When comparing the perceptron results with those of the SVM one has to consider the following: the perceptron learned from 15 drawings while the SVM learned only from 10 drawings. Secondly the perceptron was constructed with two layers of 5 cells each (inner and output layer) apart from the input cells, where the SVM has only 3 weight vectors. Thus the perceptron is expected to give more accurate results. The weight vectors for the SVM turn out to favor mainly two input values, that is (#p) and (L). It seems that the learning set permitted the SVM to train with only these very few information values, which turn out to be large scale values. For the testing set these two values are found to be sufficient for large scale aspect (b) but insufficient for small scale aspects (a) and (c ).

5. AN APPLICATION: STILL AND MOVING IMAGES OF FACES AS RECOGNISED BY INFANTS

Still and moving images of faces form a contrasting pair of types of images. For still images of faces the attention can freely flow across the whole image, whereas for moving images of faces the attention will tend to follow the movements of one part of the image for instance a moving line representing an eyebrow. To describe the flow of attention for still images of faces the kinematical model used to derive expressiveness aspect (b) seems appropriate. For moving images of faces the diffusion model for aspect (a) seems the most appropriate. Thus the two contrasting types, still and moving images, are naturally described by the large scale kinematical (L) and the small scale diffusion 1/(S) model respectively.

The attention for still and moving facial images has been studied in the case of very young infants [ref.13]. For both types, looking times of infants were measured, using schematic drawings of normal faces, and of scrambled and distorted faces. Also drawings of linear faces were used, however these drawings are very different in that they are difficult to recognize as an image of a face. The result was found that for still images, the looking time increases, while for moving faces the looking time decreases, when changing from normal to scrambled faces, for two out of
three age classes.

The drawings used for scrambled faces had larger (L) aspect and smaller 1/(S) aspect than the drawings used for normal faces. Thus for these still and moving images the looking time change is indeed described qualitatively by the kinematical (L) model and the diffusion 1/(S) model respectively. This is a confirmation of the above assumption that for still images aspect (b) and for moving images aspect (a) is most appropriate. In the case of still facial images the large scale measures of the drawing are favoured while the small scale details are eclipsed. For moving images of faces the details are appreciated most.

6.DISCUSSION

Drawings of physicists are special: they are made especially for reports during experimentation. The drawings represent, restricting ourselves to the part of the drawing with the experimental set-up as content, the experimental set-up itself, that has been built by the one who makes the drawing. Because of this motor experience, specific and individual space and time attitudes are involved. Space and time attitudes are important with respect to the content of these drawings also because of the intention of the experiment. This is totally different from, for instance, real portraits or landscapes whether drawn or photographed. The trends that were found in this study, for example (L) correlating with 1/(S), may depend on these attitudes. However it is assumed that they are present in other drawings as well.

Motor experience in general is common to everybody and it is assumed that drawings other than physics drawings will depend on it as well. Time and space attitudes depending on motor experience are a real and important source of expressiveness of drawings. Drawings made by physicists for laboratory reports are exemplary in this respect and expected to be extremely useful for studying aspects of expressiveness. Because of this they are used in this study as learning set and test set for a neural network that is set out to trace expressiveness in general. The neural network is then expected to be able to recognize extreme values of looking times , that is extremely short or extremely long looking times, as a sign of expressiveness in all kinds of line drawings

Neural network modeling is found to be a feasible and interesting way of studying expressiveness. The results obtained so far for line drawings, in this study drawings of physicists, are promising. A perceptron is found to be able to reproduce the handmade analysis of an exemplary set of line drawings. The use of a perceptron makes possible generalization of this analysis to other drawings. Among the results of the analysis is the positive correlation trend between expressiveness aspects that are both defined as looking times. Looking times for still and moving drawings or objects in general can be used to verify the substantiality of the results that are obtained by applying the neural network. The relations and differences between still and moving drawings or objects are understood here to be relying on the difference between the small and large scale expressiveness aspects. This deserves further study.

7.REFERENCES

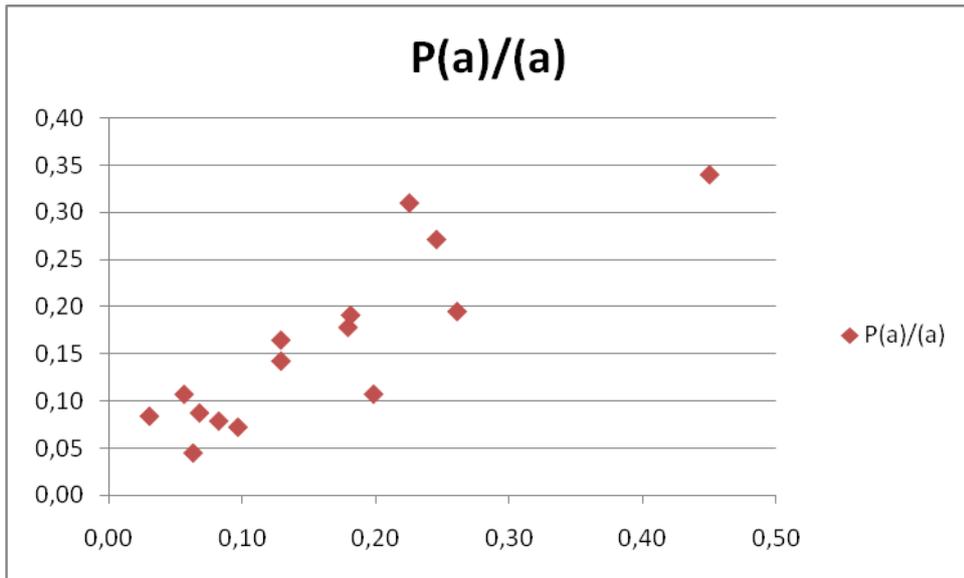
fig.1

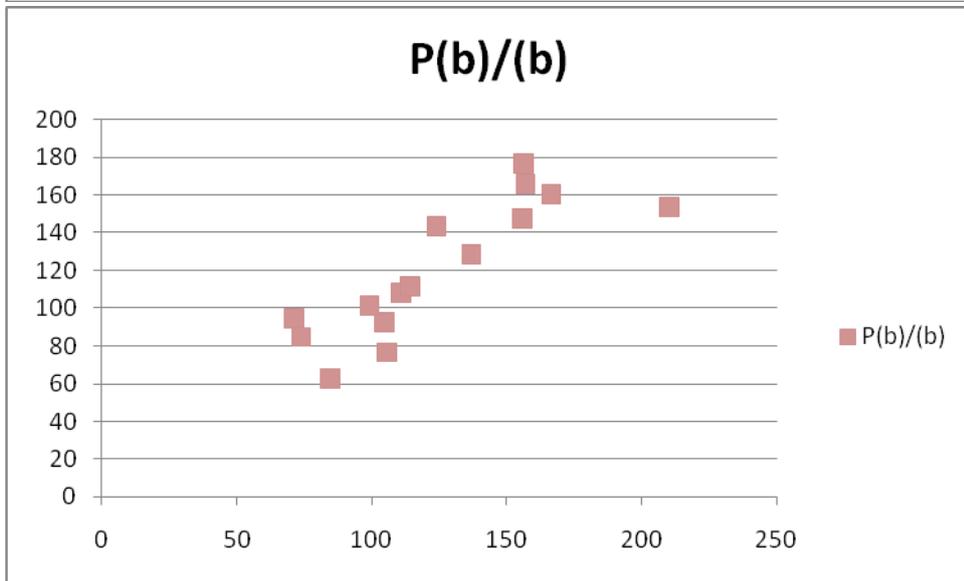
fig.2

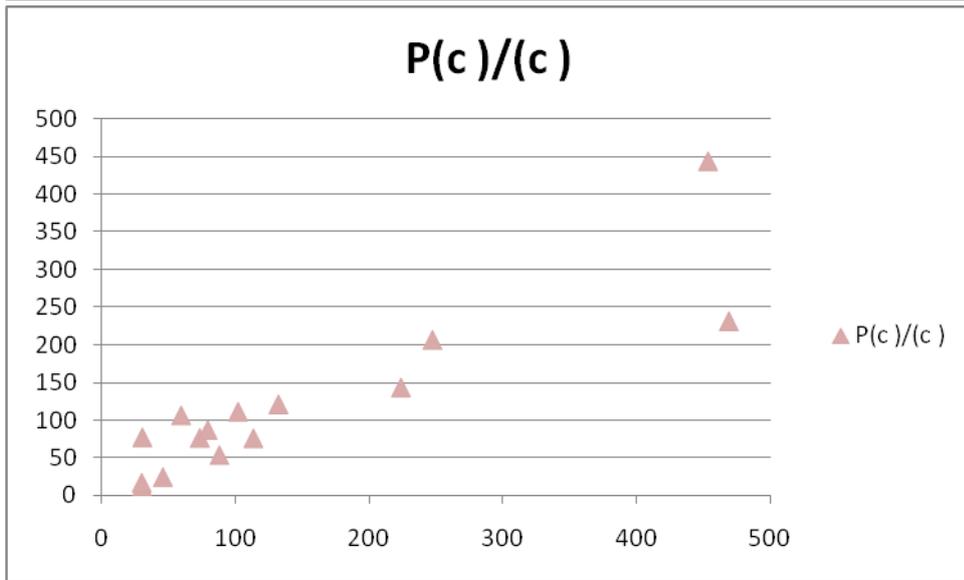
fig.3

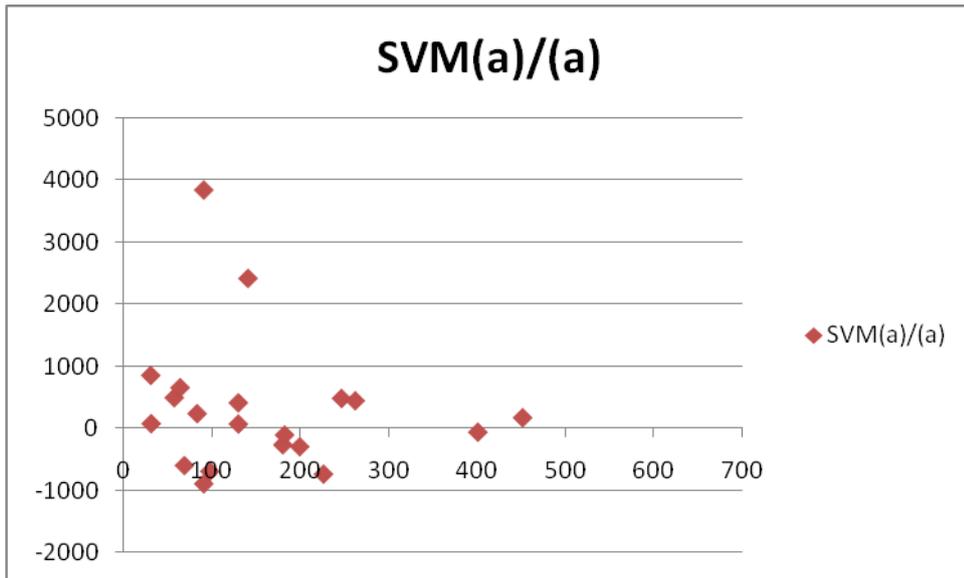

fig.4

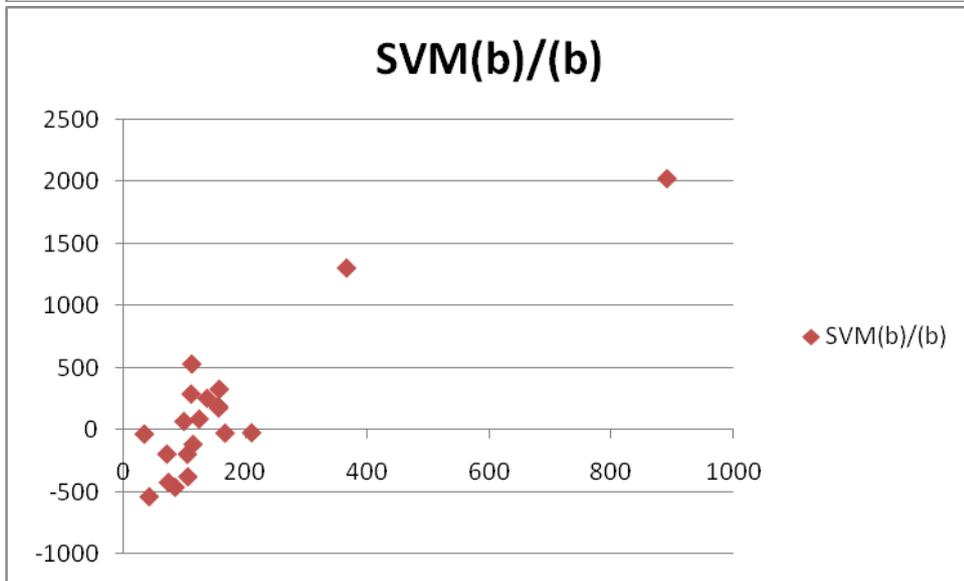

fig.5

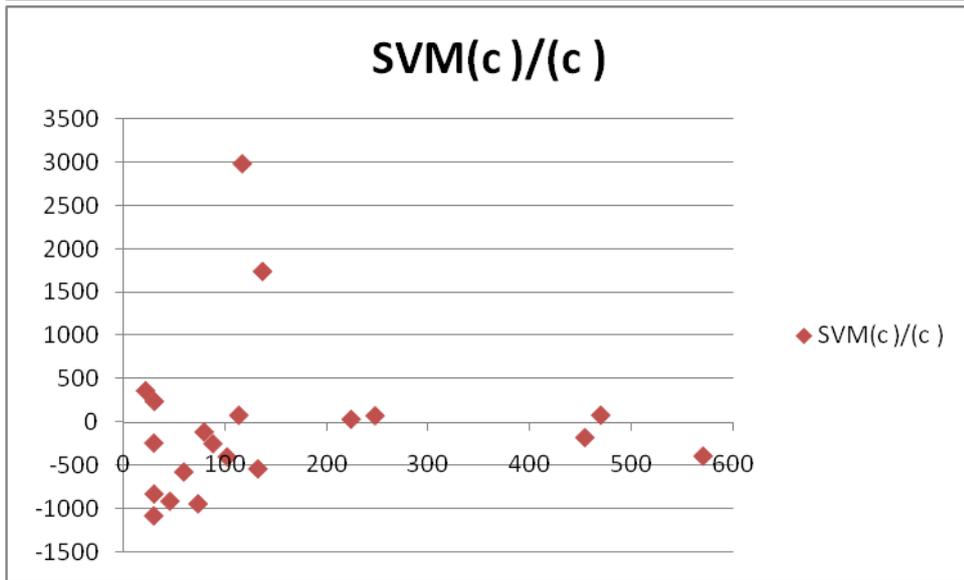

fig.6